\documentclass[twocolumn,aps,prc,superscriptaddress,showpacs]{revtex4}%\documentclass[twocolumn,showpacs,amsmath,amssymb,prl]{revtex4}
\usepackage{epsfig,epsf,graphics,psfrag}
\usepackage{times}
\usepackage{float}
\usepackage{color}
\usepackage{amsfonts,amssymb,stmaryrd,latexsym,amsmath}
\usepackage{multirow}
\usepackage{array} % for extrarowheigh
\usepackage{mathrsfs}
\usepackage{booktabs}
\usepackage{enumerate}
\usepackage{subfigure}
\usepackage{hhline}
\usepackage[hyperindex=true]{hyperref}
\usepackage{graphicx}

\usepackage{slashed}

\usepackage{verbatim}
\usepackage{esvect}
\usepackage{natbib}
\usepackage{CJK}
\usepackage{ulem}

\begin{document}
%\begin{CJK*} {GB} {gbsn}
%\begin{CJK*} {UTF8} {gbsn}

\arraycolsep1.5pt

\title{Clustering structure effect on Hanbury-Brown-Twiss correlation in $^{12}$C + $^{197}$Au collisions at 200 GeV}

\author {Junjie He}
\affiliation{Shanghai Institute of Applied Physics, Chinese Academy of Sciences, Shanghai 201800, China}
\affiliation{University of Chinese Academy of Sciences, Beijing 100049, China }

\author {Song Zhang} \thanks{Email: Corresponding author.  song\_zhang@fudan.edu.cn}
\affiliation{Key Laboratory of Nuclear Physics and Ion-beam Application (MOE), Institute of Modern Physics, Fudan University, Shanghai 200433, China}

\author {Yu-Gang Ma} \thanks{Email: Corresponding author.  mayugang@fudan.edu.cn}
\affiliation{Key Laboratory of Nuclear Physics and Ion-beam Application (MOE), Institute of Modern Physics, Fudan University, Shanghai 200433, China}
\affiliation{Shanghai Institute of Applied Physics, Chinese Academy of Sciences, Shanghai 201800, China}

\author {Jinhui Chen} %\thanks{Email: Corresponding author.  chenjinhui@fudan.edu.cn}
\affiliation{Key Laboratory of Nuclear Physics and Ion-beam Application (MOE), Institute of Modern Physics, Fudan University, Shanghai 200433, China}

\author {Chen Zhong}
\affiliation{Key Laboratory of Nuclear Physics and Ion-beam Application (MOE), Institute of Modern Physics, Fudan University, Shanghai 200433, China}

\date{\today}

\begin{abstract}

Through $^{12}$C + $^{197}$Au collisions at $\sqrt{s_{NN}} =$ 200 GeV using a multiphase transport (AMPT) model, the azimuthal angle dependences of the Hanbury Brown-Twiss (HBT) radii relative to the second- and third-order participant plane from $\pi$-$\pi$ correlations are discussed.
Three initial geometric configurations of $^{12}$C, namely three-$\alpha$-cluster triangle, three-$\alpha$-cluster chain and Woods-Saxon distribution of nucleons, are taken into account, and their effects on the correlations are investigated.
The ratio of the third- to the second-order HBT radii $R_{o(s),3}^2/R_{o(s),2}^2$ is shown to be a clear probe for three configurations.
In addition, this work presents the hadronic rescattering time evolution of the azimuthally dependent HBT radii.
From the present study, one can learn that the HBT correlation from identical particles at freeze-out is able to provide the information of different initial configurations as collective flow proposed before.

\end{abstract}

\pacs{25.75.Gz, 21.65.Qr, 25.75.Nq}

 \maketitle

\section{Introduction}
\label{SecIntro}

Clustering, which could exist extensively in the ground and excited states of $N = Z$ nuclei and nuclei far from the $\beta$ stability line, is a fundamental physics aspect in light nuclei.
A great number of detailed studies have been focused on clustering for more than 50 years~\cite{greiner1995nuclear,von2006nuclear}.
$\alpha$ cluster, proposed by Gamow~\cite{gamow1931constitution}, is the most basic object and plays a critical role in light nuclei clustering due to its high stability and the strong repulsive $\alpha$-$\alpha$ interaction~\cite{von2006nuclear}.
In order to explain the excited states of $\alpha$-conjugate nuclei (e.g., $^{12}$C and $^{16}$O), which cannot be described by the shell model, Ikeda and his collaborators proposed an $\alpha$-clustering diagram in 1968~\cite{ikeda1968systematic}, suggesting that these states could be associated with configurations composed of $\alpha$ particles and predicting that cluster structures could be mainly found close to cluster decay thresholds, which was verified experimentally later on.
In general, the $\alpha$-conjugate light nuclei are expected to experience a transition in which nucleons condense into $\alpha$ particles~\cite{brink1973alpha}, and then they are either in an $\alpha$-gas or an $\alpha$-particle Bose-Einstein condensate~\cite{yamada2012nuclear} as the density decreases, which will considerably change the nuclear equation of state~\cite{horowitz2006cluster,natowitz2010symmetry,qin2012laboratory,Pias-nst}.

In the last several decades, numerous complex cluster structures composed of $\alpha$ particles in $\alpha$-conjugate nuclei have been found, among which the $^{12}$C nucleus is of particular interest.
For instance, the Hoyle state in $^{12}$C at 7.65 MeV, whose existence is essential for the nucleosynthesis of carbon via the triple-$\alpha$ process~\cite{hoyle1954nuclear}, is believed to be formed out of $\alpha$ clusters according to the calculations from different models, however, its specific configuration or property of clusters is still arguable.
Configuration of $\alpha$ clusters is a key problem to understand the clustering in light nuclei~\cite{von2006nuclear,YeYL-nst,Aygun-nst}, therefore, further research and new methods are certainly of great interest.

Recently, He and Ma $et$ $al.$ found that the giant dipole resonance (GDR) spectrum was highly fragmented into several apparent peaks due to the $\alpha$-clustering structure, which demonstrated an effective probe to discriminate the different $\alpha$-clustering configurations in light nuclei~\cite{he2014giant,he2016dipole}.
Also, the collective flows at Fermi energy~\cite{Guo,Guo2}, the neutron and proton emission from photon induced reactions in quasi-deuteron regime~\cite{Huang1,Huang2} can also be taken as useful probes of clustering structure from an-extended-quantum-molecular-dynamics-model calculations.
While, Broniowski $et$ $al.$ proposed that through relativistic heavy-ion collision, such as $^{12}$C + $^{208}$Pb collision, collective flow can be the signature of $\alpha$-clustering in light nuclei in their ground state, which offers a possibility of studying low-energy nuclear structure phenomena with observables in relativistic heavy-ion collisions~\cite{broniowski2014signatures,bozek2014alpha}.
In our previous work, the initial geometry effect on elliptic and triangular flows as well as global rotation has been studied in $^{12}$C + $^{197}$Au collisions at $\sqrt{s_{NN}} =$ 200 GeV and 10 GeV~\cite{Zhang:2017xda,Zhang:2018,XuZW-nst}.
It is expected that the shape of the created fireball in the transverse plane reflects the initial geometry of the light nucleus, because at ultrahigh collision energies, where nucleon-nucleon inelastic collisions produce abundant particles, the interaction time is short enough to prevent the much slower nuclear excitations.
In this work, we further argue the feasibility of another observable, namely the Hanbury Brown-Twiss (HBT) radii of $\pi$-$\pi$ correlation as an effective probe of $\alpha$-clustering in light nuclei.
As a specific example we present a study of $^{12}$C + $^{197}$Au collisions at $\sqrt{s_{NN}} =$ 200 GeV, and the configuration of $^{12}$C includes triangular and linear 3-$\alpha$-cluster structures as well as the spherical nucleon distribution.

In ultra-relativistic heavy-ion collisions at BNL-RHIC and CERN-LHC, the quark-gluon plasma (QGP) is believed to be created~\cite{arsene2005quark,back2005phobos,adams2005experimental,adcox2005formation,chen2018antinuclei,andronic2018decoding,luo2017search,Jin2018}.
Studying the properties of QGP is one of the main goals of these experiments.
Detailed experimentally driven understanding of the freeze-out configuration is the crucial first step to learn the system evolution and the physics of hot dense quark-gluon matter.
For example, a detailed understanding of the space-time geometry and dynamics of fireball is required.
Because of the short lifetime and extension of the source, two-particle intensity interferometry, or the so-called HBT method, provides a unique way to obtain direct information about the space-time structure from the measured particle momenta.

Two-particle intensity interferometry was proposed and developed by Hanbury Brown and Twiss in the 1950s, who used two-photon intensity interferometry to measure the angular diameter of Sirius and other astronomical objects~\cite{brown1956correlation,brown1956test}.
In particle physics, this effect was discovered by Goldhaber $et$ $al.$ in 1960 when they studied the angular correlations between identical pions in proton-antiproton collisions.
They observed an enhancement of two-pion correlation function at small relative momenta and explained this effect (GGLP effect) in terms of the Bose-Einstein statistics~\cite{goldhaber1960influence}.
Later, it was gradually realized that the correlations of identical particles were sensitive not only to the geometry of the source, but also to its lifetime~\cite{Kopylov:1972qw,Shuryak:1972kq}, and even that the pair momentum dependence of the correlations contained information about the collision dynamics~\cite{pratt1984pion,pratt1986coherence}.
Other important progress included a more detailed analysis of the effect of final state interactions~\cite{koonin1977proton,gyulassy1981coulomb,bowler1988extended}, the development of a parametrization considering the longitudinal expansion of the source~\cite{pratt1986coherence} and the implementation of the HBT effect in prescriptions for event generator studies~\cite{zajc1987monte}.
Throughout the last decades, the HBT method has been extensively applied in heavy-ion collisions, see e.g.~\cite{Lisa,Nature-pbar,Ma-22Mg,Fang-22Mg,ZhangZQ,ZhangWN}.

On the other hand, in order to understand the extensive experimental results of different observables, many theoretical models were introduced, ranging from thermal models~\cite{braun1995thermal,cleymans1998unified,becattini2001features}, hydrodynamic models~\cite{huovinen2001radial,betz2009universality,schenke2011elliptic} to transport models~\cite{xu2005thermalization,lin2005multiphase,cassing2009parton}.
Simulations of the space-time evolution of relativistic heavy-ion collisions have been performed extensively with hydrodynamic models, transport models, and hybrid models that combine a hydrodynamic model with a transport model~\cite{petersen2008fully,werner2010event,song2011200,song2012erratum,Song}.
Many observables such as azimuthal anisotropies have been successfully described by all three types of models in spite of the different physics foundations and assumptions.

In this paper, we simulate the high-energy $^{12}$C + $^{197}$Au collisions and discuss the influences of initial geometry on the observables at freeze-out stage using a multiphase transport (AMPT) model~\cite{lin2005multiphase}.
The article is arranged as follows: in Sect.~\ref{SecAMPT} the AMPT Model and algorithm introduction for $^{12}$C initialization are described.
HBT correlation function and participant plane are introduced in Sect.~\ref{SecHBT}.
Results and discussion for the effect of cluster configuration on the second- and third-order HBT radii as well as eccentricities are then presented in Sect.~\ref{SecResults}.
Finally, a summary is given in Sect.~\ref{SecSummary}.

\section{The AMPT Model and algorithm introduction}
\label{SecAMPT}

The AMPT model is a successful model for studying reaction dynamics in relativistic heavy-ion collisions, consisting of four main components: the initial conditions, partonic interactions, hadronization and hadronic interactions.
There are two versions of the AMPT model: the default version and the string melting version.

The phase-space distributions of minijet partons and soft string excitations are initialized by the heavy ion jet interaction generator (HIJING) model~\cite{gyulassy1994hijing}.
Scatterings among partons are modeled by Zhang's parton cascade (ZPC)~\cite{zhang1998zpc}.
In the default version, minijet partons are recombined with the parent strings to form new excited strings after partonic interactions.
Hadronization of these strings is described by the Lund string fragmentation model~\cite{sjostrand1994high}.
In the version of string melting, both excited strings and minijet partons are decomposed into partons, whose evolution in time and space is modeled by the ZPC model.
After partons stop interacting with each other, they are converted to hadrons via a simple quark coalescence model.
In both versions, the interactions of the subsequent hadrons are described by a hadronic cascade, which is based on a relativistic transport (ART) model~\cite{li1995formation}.

The default AMPT model is able to give a reasonable description of the rapidity distributions and $p_T$ spectra in heavy-ion collisions from CERN Super Proton Synchrotron (SPS) to RHIC energies.
However, it underestimates the elliptic flow observed at RHIC, and the reason is that most of the energy produced in the overlap volume of heavy-ion collisions is in hadronic strings and thus not included in the parton cascade in the model~\cite{lin2002partonic}.
The string melting AMPT model is employed in this calculation and the hadronic rescattering time is set as different time, namely 30 fm/$c$, 7 fm/$c$ and zero (with resonance decay but without hadronic rescattering) for the small colliding system relative to $^{197}$Au+$^{197}$Au collision.

Several theoretical predictions have been made on $\alpha$-clustering configuration of $^{12}$C.
For an example, a triangle-like configuration is predicted around the ground state by Fermionic molecular dynamics~\cite{chernykh2007structure}, antisymmetrized molecular dynamics~\cite{Kanada-Enyo:2012yif}, and covariant density functional theory~\cite{liu2012alpha}, which is experimentally supported~\cite{Marin-Lambarri:2014zxa}.
A three-$\alpha$ linear-chain configuration is predicted as an excited state in time-dependent Hartree-Fock theory~\cite{Umar:2010ck} and other different approaches~\cite{Morinaga1966On}.
Together with the situation of Woods-Saxon (W-S) distribution, we initialize the configurations of $^{12}$C by these three cases in HIJING process.

\begin{figure*}[htbp]
\centering
\includegraphics[width=1\textwidth]{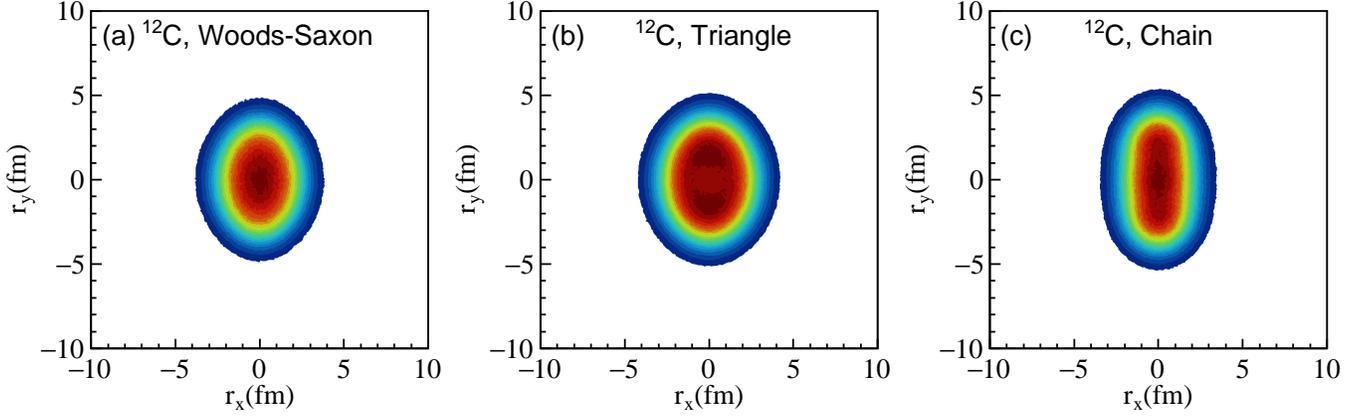}
\caption{Participant distributions of $^{12}$C + $^{197}$Au central collisions with different initial configurations of $^{12}$C.}
\label{density}
\end{figure*}

Fig.~\ref{density} shows the participant nucleon distributions of $^{12}$C + $^{197}$Au central collisions for these three cases.
The initial nucleon distribution in $^{12}$C is configured as (a) nucleons in Woods-Saxon distribution from the HIJING model~\cite{gyulassy1994hijing}, (b) three $\alpha$ clusters in triangle structure, and (c) three $\alpha$ clusters in chain structure.
The distribution of the radial center of the $\alpha$ clusters in $^{12}$C is assumed to be a Gaussian function, $e^{-0.5\left(\frac{r-r_c}{\sigma_{r_c}}\right)^{2}}$, here $r_c$ is the average radial center of an $\alpha$ cluster and $\sigma_{r_c}$ is the width of the distribution.
And the nucleons inside each $\alpha$ cluster are given by the three-parameter Woods-Saxon distribution in HIJING.
The parameter $r_c$ is set to match the measured value of the rms nuclear charge radius of $^{12}$C.
For triangle structure, $r_c =$ 1.79 fm and $\sigma_{r_c} =$ 0.1 fm.
For chain structure, $r_c =$ 2.19 fm, $\sigma_{r_c} =$ 0.1 fm for two $\alpha$ clusters and the other one is at the center of $^{12}$C.
Once the radial center of the $\alpha$ cluster is determined, the centers of the three clusters will be placed in an equilateral triangle for triangle structure or in a line for chain structure.
From the parameters and initialization, we obtain that the rms radius $\langle R(\text{WS}) \rangle \approx$ 2.46 fm for Woods-Saxon distribution.
For the other two configurations, $\langle R(\text{Chain}) \rangle \approx$ 2.47 fm and $\langle R(\text{Triangle}) \rangle \approx$ 2.47 fm.
These values are all consistent with experimental data 2.47 fm~\cite{ANGELI201369}.

\section{HBT Correlation Function}
\label{SecHBT}

The identical two-particle HBT correlation function without final state interactions, defined as the ratio of the two-particle probability divided by the product of the single-particle probabilities, can be written as~\cite{Heinz:1999rw}
\begin{equation}
C(\vec{q},\vec{K}) = 1 \pm \left| \frac{\int d^4x e^{i \vec{q} \cdot (\vec{x} - \vec{\beta}t)} S(x,K)}{\int d^4x S(x,K)}\right|^2,
\end{equation}
where $K$ is the average momentum of two particles, $K = \frac{1}{2}(p_1+p_2)$, $q$ is the relative momentum between two particles, $q = p_1 - p_2$, and $S(x,K)$ is the emission function.
The positive sign is for bosons and the negative one is for fermions.
With the mass-shell constraint $q \cdot K = 0$, the temporal component $q_0$ in the four-dimensional Fourier transform is substituted with $\vec{\beta} = \frac{\vec{K}}{K_0}$.
And, both smoothness approximation and on-shell approximation are used.
With the assumption of Gaussian source, the correlation function will also take a Gaussian form.
Further, in order to obtain information of the source in the beam and transverse directions, Cartesian or Bertsch-Pratt parametrization is taken.
In this parametrization, the relative momentum vector of the pair $\vec{q}$ is decomposed into a longitudinal direction along the beam axis, $q_l$, an outward direction parallel to the pair transverse momentum, $q_o$, and a sideward direction perpendicular to those two, $q_s$.
Therefore, the correlation function can be expressed as~\cite{pratt1984pion,Podgoretsky:1982xu,Bertsch:1988db}
\begin{equation}
C(\vec{q},\vec{K}) = 1 + \lambda(\vec{K})\mathrm{exp}(-\sum_{i,j = o,s,l} R_{ij}^2(\vec{K})q_i q_j).
\label{CF}
\end{equation}
where $\lambda$ is coherence parameter.
It should be unity for a fully chaotic source and smaller than unity for a partially coherent source, although in practice many other effects, e.g. particle misidentification, resonance decay contributions and final state Coulomb interactions, can decrease the measured $\lambda$ significantly.
After taking the notation,
\begin{equation}
\langle \xi \rangle = \frac{\int d^4x \xi S(x,K)}{\int d^4x S(x,K)},
\label{average}
\end{equation}
the explicit azimuthal angle dependence of the HBT radii can be seen from~\cite{Wiedemann:1997cr},
\begin{equation}
  \begin{split}
R_s^2(K_T,\Phi,Y) = &\langle (y \cos \Phi - x \sin \Phi)^2 \rangle\\
&- \langle y \cos \Phi -x \sin \Phi \rangle^2,
  \end{split}
\label{Rs}
\end{equation}
\begin{equation}
  \begin{split}
R_o^2(K_T,\Phi,Y) = &\langle (x \cos \Phi + y \sin \Phi - \beta_\perp t)^2 \rangle\\
&- \langle x \cos \Phi + y \sin \Phi - \beta_\perp t \rangle^2,
  \end{split}
\label{Ro}
\end{equation}
where $Y = \frac{1}{2}\ln(\frac{E_K+K_l}{E_K-K_l})$, $\beta_\perp = K_T/K_0$, 
and $K_0 \approx E_K$.
The other terms of the HBT radii will not be studied in this work and the expression can also be found in reference~\cite{Wiedemann:1997cr}.

In heavy-ion collisions, the direction of impact parameter $\vec{b}$ is always random and along with the line between the centers of target and projectile.
The so-called reaction plane is formed by $\vec{b}$ and the beam direction, which results in random distribution of reaction plane angle.
Hence, the observables sensitive to reaction plane angle should be corrected to that.
Participant plane is always considered as a reasonable approximation to reaction plane, and the participant plane angle $\Psi_n\{PP\}$ can be defined by~\cite{Alver:2010gr, Voloshin:2007pc, Lacey:2010hw},
\begin{equation}
\Psi_n\{PP\} = \frac{\mathrm{atan2}\left(\frac{\left<r_{part}^2\sin\left(n\phi_{part}\right)\right>}{\left<r_{part}^2\cos\left(n\phi_{part}\right)\right>}\right)+\pi}{n},
\label{PartPlanDef}
\end{equation}
where $\Psi_n\{PP\}$ is the $n$th-order participant plane angle, $r_{part}$ and $\phi_{part}$ are radial coordinate and azimuthal angle of participants in the collision zone in the initial state, and the average $\left<\cdots\right>$ denotes density weighting.
The system will be rotated to the participant plane direction in transverse plane.
In references~\cite{Plumberg:2013nga,Bozek:2014hwa}, the azimuthal angle dependence of the HBT radii was investigated and the effects of geometric deformation and anisotropic flow were discussed therein.
And in experiments, RHIC-STAR~\cite{Adams:2004yc,Adams:2003ra}, RHIC-PHENIX~\cite{Adare:2015bcj} and CERN-ALICE~\cite{Acharya:2018dpu} reported the measurements of $\pi$-$\pi$ correlation relative to the second and third harmonic event plane, respectively.
In this calculation we will focus on the azimuthal angle dependence of the HBT radii relative to the second- and third-order participant plane and investigate the initial geometry effect.

\section{Results and discussion}
\label{SecResults}

$\pi^{\pm}$ meson, which is the most produced particle in relativistic heavy-ion collision, is taken for the HBT correlation calculation.
The transverse momentum spectrum of pions was calculated and basically insensitive to the initial condition used (Woods-Saxon distribution, chain structure and triangle structure of $^{12}$C), which was consistent with our previous work~\cite{Zhang:2017xda} that the multiplicity was not so dependent on the initial configuration.
And then it will be an interesting question how the initial distribution affects the differential distribution of particles in momentum and coordinate space at the final stage of the collision.
Next we focused on the HBT radii from Eqs.~(\ref{Rs}) and (\ref{Ro}) relative to the second- and third-order participant plane angle.

\begin{figure*}[htbp]
\centering
\includegraphics[width=1\textwidth]{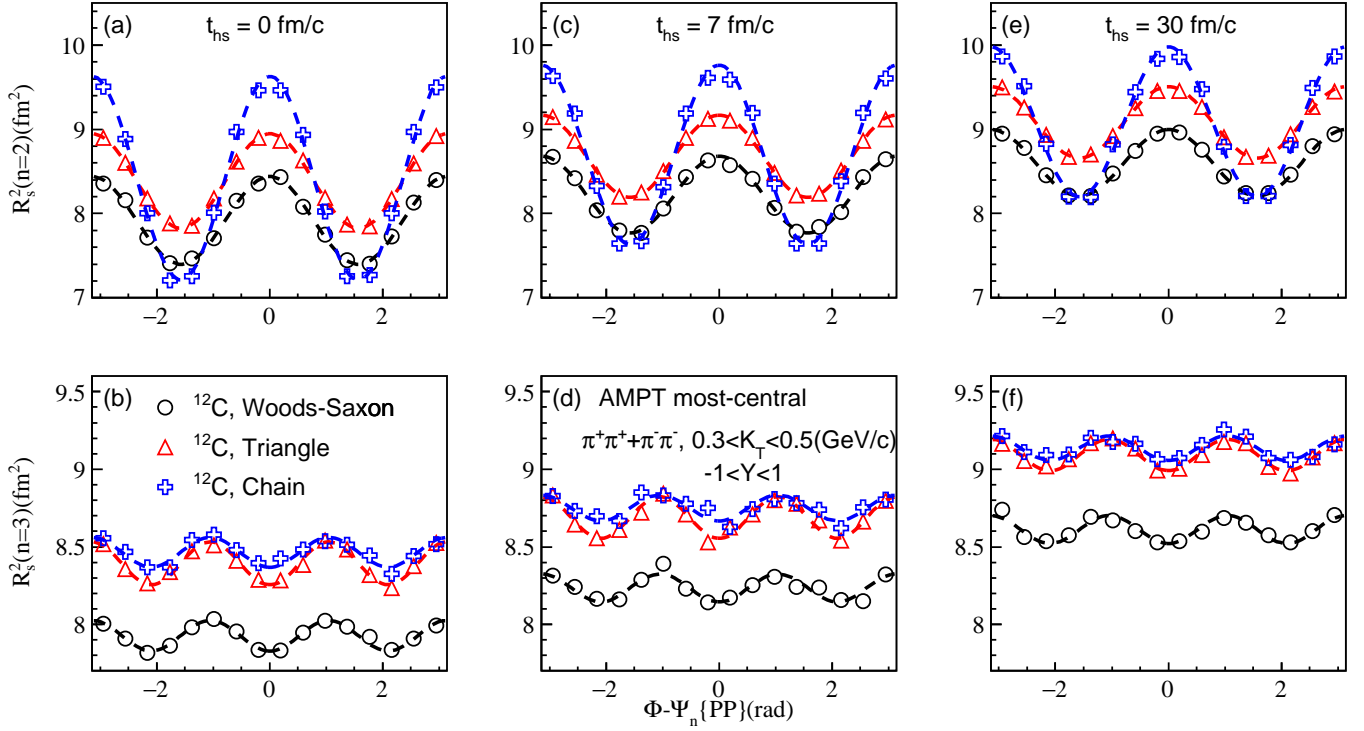}
\caption{Azimuthal angle dependence of the HBT radii $R_s^2$ relative to the second (top row $n=2$) and third (bottom row $n=3$) harmonic participant plane $\Psi_n\{PP\}$ from $\pi$-$\pi$ correlation with $K_T$ from 0.3 GeV/$c$ to 0.5 GeV/$c$ for the most central $^{12}$C + $^{197}$Au collisions.
The results from left column to right column correspond to the hadronic rescattering time $t_{hs} =$ 0 fm/$c$, $t_{hs} =$ 7 fm/$c$ and $t_{hs} =$ 30 fm/$c$, respectively.
The dots represent the calculated results and the curves are the fits by Eq.~(\ref{fits}).}
\label{HBT-pi-Rs-1}
\end{figure*}

\begin{figure*}[htbp]
\centering
\includegraphics[width=1\textwidth]{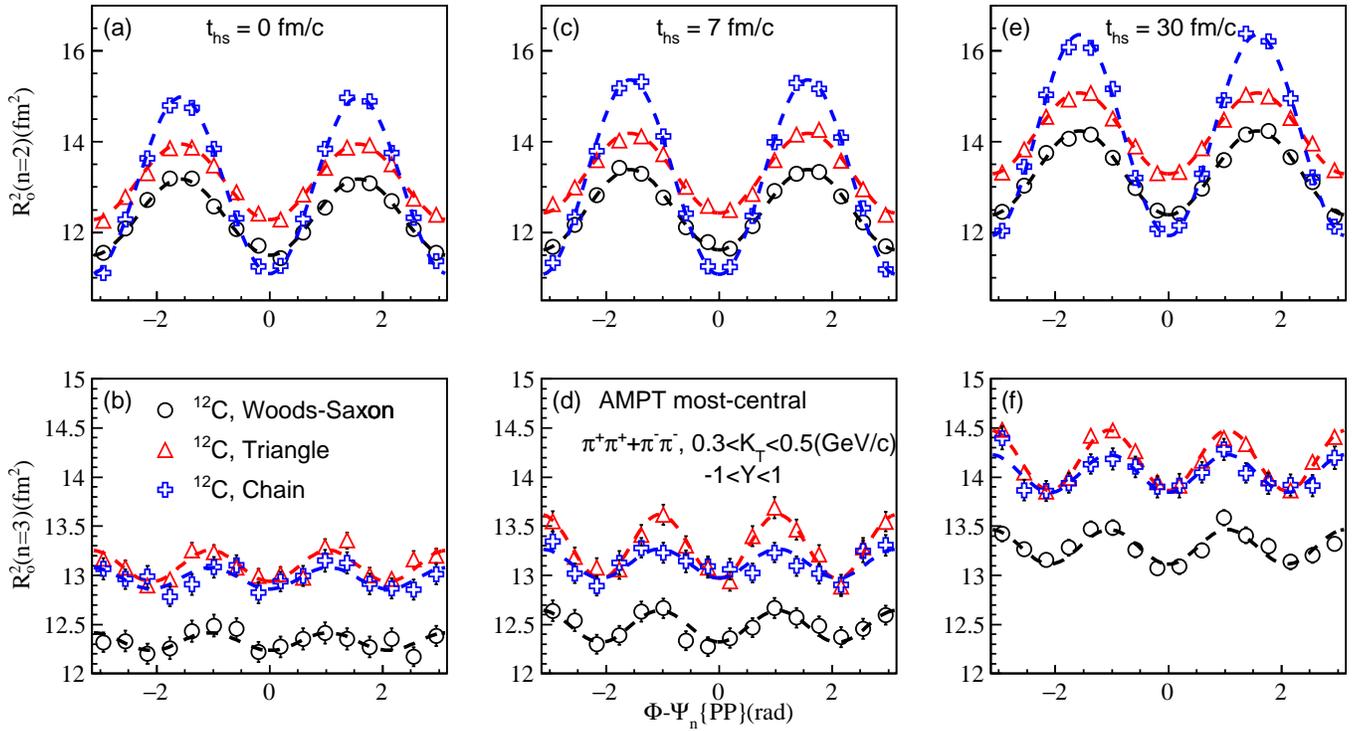}
\caption{Same as Fig.~\ref{HBT-pi-Rs-1} but for $R_o^2$.}
\label{HBT-pi-Ro-1}
\end{figure*}

\begin{figure*}[htbp]
\centering
\includegraphics[width=1\textwidth]{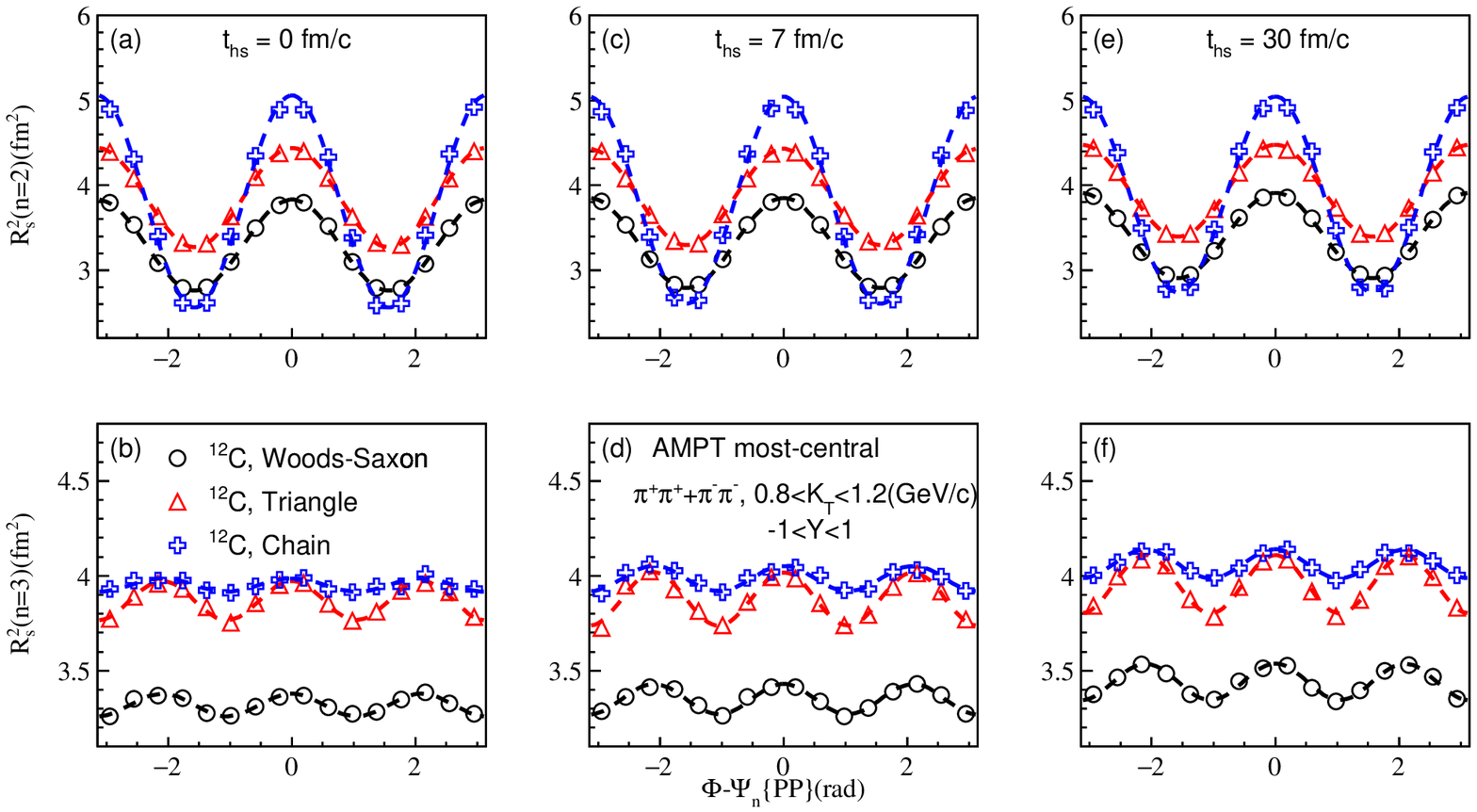}
\caption{Same as Fig.~\ref{HBT-pi-Rs-1} but with $K_T$ from 0.8 GeV/$c$ to 1.2 GeV/$c$.}
\label{HBT-pi-Rs-2}
\end{figure*}

\begin{figure*}[htbp]
\centering
\includegraphics[width=1\textwidth]{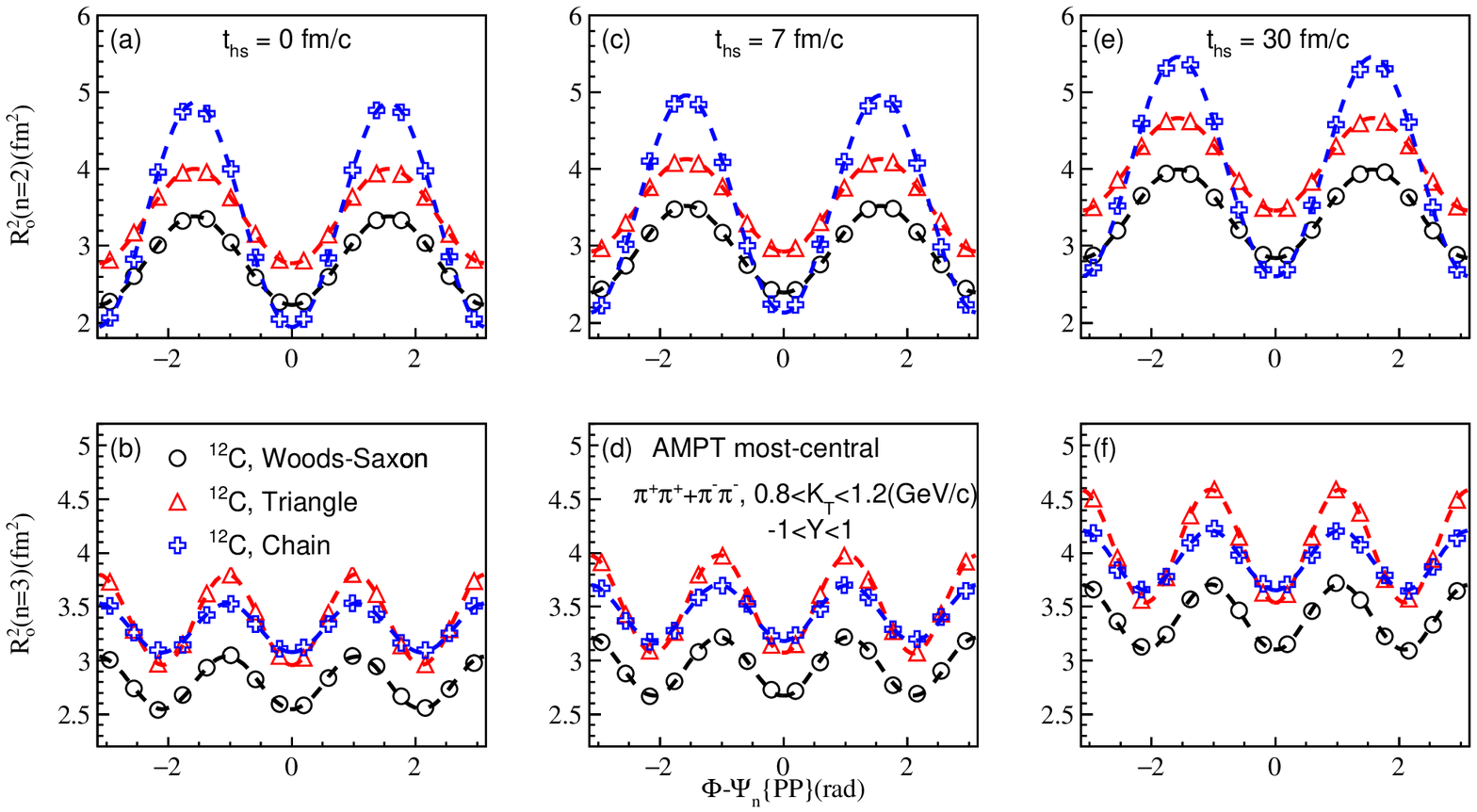}
\caption{Same as Fig.~\ref{HBT-pi-Ro-1} but with $K_T$ from 0.8 GeV/$c$ to 1.2 GeV/$c$.}
\label{HBT-pi-Ro-2}
\end{figure*}

Figures~\ref{HBT-pi-Rs-1},~\ref{HBT-pi-Ro-1},~\ref{HBT-pi-Rs-2},~\ref{HBT-pi-Ro-2} display the azimuthal angle dependence of the HBT radii, including $R_s^2$ and $R_o^2$ as functions of $\Phi-\Psi_2\{PP\}$ or $\Phi-\Psi_3\{PP\}$, from $\pi$-$\pi$ correlation with three different initial structures of $^{12}$C as well as different hadronic rescattering time.
The calculations are performed with rapidity from -1 to 1 and two $K_T$ bins (0.3-0.5 GeV/$c$ and 0.8-1.2 GeV/$c$) in the most central $^{12}$C + $^{197}$Au collisions at $\sqrt{s_{NN}} =$ 200 GeV.
$t_{hs} =$ 0 fm/$c$ means hadrons with resonance decay but without hadronic rescattering and $\Phi$ is the azimuthal angle of pion.
From the hadronic rescattering time $t_{hs}$ evolution, it can be seen that the azimuthal angle dependence of the HBT radii keeps the similar dependent structure and only the amplitude has minor changes.
In other words, the hadronic rescattering does not destroy the azimuthal angle dependence of the HBT radii from the early stage of the fireball.

For the azimuthal angle dependence of the HBT radii relative to the second-order participant plane angle $\Psi_2\{PP\}$, shown in panel (a), (c), (e) in Figs.~\ref{HBT-pi-Rs-1},~\ref{HBT-pi-Ro-1},~\ref{HBT-pi-Rs-2} and~\ref{HBT-pi-Ro-2}, $R_s^2$(n=2) and $R_o^2$(n=2) present $\cos[2(\Phi - \Psi_2\{PP\})]$ oscillation.
$R_s^2$(n=2) shows a peak and valleys around $\Phi-\Psi_2\{PP\} =$ 0 and $\pm \pi/2$, respectively, but it is reverse for $R_o^2$(n=2).
This azimuthal angle dependence of the HBT radii relative to $\Psi_2\{PP\}$ is similar to experimental results in Au+Au collisions at $\sqrt{s_{NN}} =$ 200 GeV~\cite{Adams:2004yc,Adams:2003ra,Adare:2015bcj}.
The three cases of initial $^{12}$C configuration present the same oscillation pattern for the HBT radii as a function of azimuthal angle but with different amplitudes.

For the azimuthal angle dependence of the HBT radii relative to the third-order participant plane angle $\Psi_3\{PP\}$, shown in panel (b), (d), (f) in Figs.~\ref{HBT-pi-Rs-1},~\ref{HBT-pi-Ro-1},~\ref{HBT-pi-Rs-2} and~\ref{HBT-pi-Ro-2}, $R_s^2$(n=3) and $R_o^2$(n=3) oscillate in $\cos[3(\Phi - \Psi_3\{PP\})]$ mode.
$R_o^2$(n=3) shows valleys around $\Phi-\Psi_3\{PP\} =$ 0 and $\pm 2\pi/3$ for both $K_T$ bins.
However, $R_s^2$(n=3) has a different behavior.
With high $K_T$ bin, $R_s^2$(n=3) shows a peak at 0 as $R_s^2$(n=2), while it is reverse with low $K_T$ bin.
Similar behavior was observed in a toy model simulation, and was explained by flow anisotropy dominated source~\cite{Plumberg:2013nga}.

From the above results, it can be seen that the initial geometric distribution from the nuclear structure can be observed from the azimuthal angle dependence of the HBT radii relative to the second- and third-order participant plane.
In order to quantitatively describe the anisotropy of the above HBT radii and discuss the effect from exotic nuclear structure, we perform a Fourier expansion in azimuthal angle for the HBT radii which can be found in references~\cite{Plumberg:2013nga,Bozek:2014hwa,Wiedemann:1997cr,retiere2004observable},
\begin{equation}
  \begin{split}
R_{s}^2(\Phi - \Psi_n) = &R_{s,0}^2 + 2R_{s,n}^2\cos[n(\Phi - \Psi_{n})],\\
R_{o}^2(\Phi - \Psi_n) = &R_{o,0}^2 + 2R_{o,n}^2\cos[n(\Phi - \Psi_{n})], n=2,3
  \end{split}
  \label{fits}
\end{equation}
where $R_{s(o),n}^2$ is the so-called zeroth-, second- (n=2) and third- (n=3) order HBT radii respectively.

\begin{table}[htbp]
\renewcommand\arraystretch{1.5}
\caption{ \label{tab:HBTRFitPars-1} $R_{s(o),n}^2$ extracted by Eq.~(\ref{fits}) from azimuthal angle dependence of the HBT radii relative to the second- and third-order participant plane with $K_T$ from 0.3 GeV/$c$ to 0.5 GeV/$c$ for different configurations of $^{12}$C.}
\begin{ruledtabular}
\begin{tabular}{llll}
$t_{hs}$ (fm/$c$) & 0 & 7 & 30\\\hline
Woods-Saxon & & &\\\hline
$R_{s,0}^2$ & 7.92 & 8.23 & 8.60\\
$R_{s,2}^2$ & 0.260 & 0.228 & 0.197\\
$R_{s,3}^2$ & -0.0490 & -0.0448 & -0.0449\\
$R_{o,0}^2$ & 12.3 & 12.5 & 13.3\\
$R_{o,2}^2$ & -0.418 & -0.441 & -0.462\\
$R_{o,3}^2$ & -0.0447 & -0.0806 & -0.0881\\\hline
Triangle & & &\\\hline
$R_{s,0}^2$ & 8.39 & 8.68 & 9.08\\
$R_{s,2}^2$ & 0.280 & 0.245 & 0.212\\
$R_{s,3}^2$ & -0.0693 & -0.0670 & -0.0508\\
$R_{o,0}^2$ & 13.1 & 13.3 & 14.2\\
$R_{o,2}^2$ & -0.417 & -0.439 & -0.447\\
$R_{o,3}^2$ & -0.0790 &-0.163 & -0.154\\\hline
Chain & & &\\\hline
$R_{s,0}^2$ & 8.42 & 8.70 & 9.09\\
$R_{s,2}^2$ & 0.605 & 0.532 & 0.445\\
$R_{s,3}^2$ & -0.0480 & -0.0414 & -0.0393\\
$R_{o,0}^2$ & 13.0 & 13.2 & 14.1\\
$R_{o,2}^2$ & -0.973 & -1.07 & -1.11\\
$R_{o,3}^2$ & -0.0547 & -0.0740 & -0.0960
\end{tabular}
\end{ruledtabular}
\end{table}

\begin{table}[htbp]
\renewcommand\arraystretch{1.5}
\caption{ \label{tab:HBTRFitPars-2} Same as Table~\ref{tab:HBTRFitPars-1} but with $K_T$ from 0.8 GeV/$c$ to 1.2 GeV/$c$.}
\begin{ruledtabular}
\begin{tabular}{llll}
$t_{hs}$ (fm/$c$) & 0 & 7 & 30\\\hline
Woods-Saxon & & &\\\hline
$R_{s,0}^2$ & 3.29 & 3.32 & 3.41\\
$R_{s,2}^2$ & 0.269 & 0.266 & 0.251\\
$R_{s,3}^2$ & 0.0296 & 0.0396 & 0.0478\\
$R_{o,0}^2$ & 2.81 & 2.96 & 3.42\\
$R_{o,2}^2$ & -0.288 & -0.281 & -0.289\\
$R_{o,3}^2$ & -0.123 & -0.136 & -0.151\\\hline
Triangle & & &\\\hline
$R_{s,0}^2$ & 3.85 & 3.86 & 3.94\\
$R_{s,2}^2$ & 0.292 & 0.284 & 0.272\\
$R_{s,3}^2$ & 0.0514 & 0.0701 & 0.0767\\
$R_{o,0}^2$ & 3.39 & 3.53 & 4.06\\
$R_{o,2}^2$ & -0.309 & -0.300 & -0.301\\
$R_{o,3}^2$ & -0.211 & -0.228 & -0.262\\\hline
Chain & & &\\\hline
$R_{s,0}^2$ & 3.81 & 3.82 & 3.89\\
$R_{s,2}^2$ & 0.625 & 0.610 & 0.577\\
$R_{s,3}^2$ & 0.0175 & 0.0333 & 0.0379\\
$R_{o,0}^2$ & 3.41 & 3.54 & 4.03\\
$R_{o,2}^2$ & -0.732 & -0.708 & -0.715\\
$R_{o,3}^2$ & -0.114 & -0.131 & -0.139
\end{tabular}
\end{ruledtabular}
\end{table}

\begin{figure*}[htbp]
\centering
\includegraphics[width=1\textwidth]{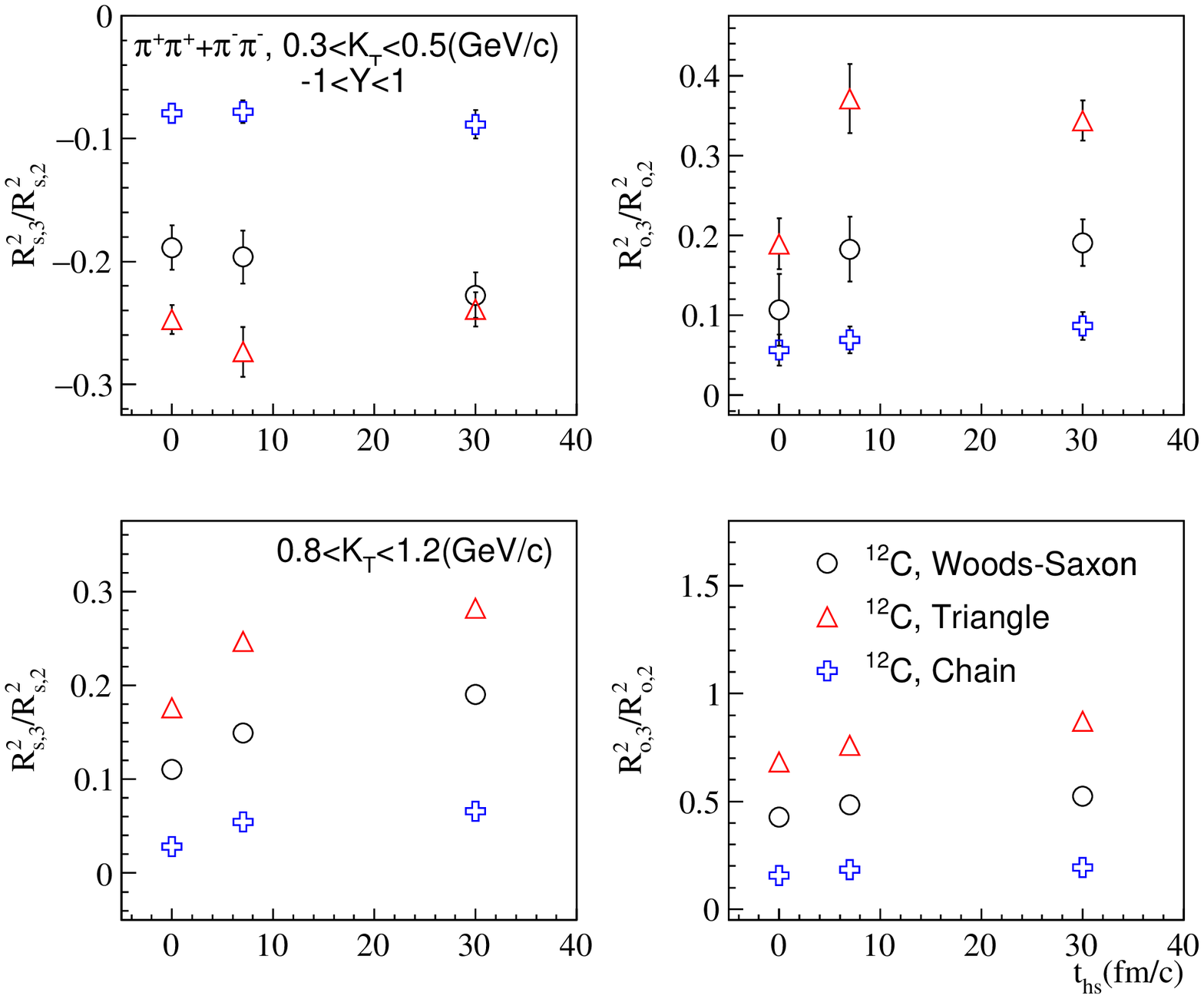}
\caption{The time evolution of the ratio of the third- to the second-order HBT radii $R_{s(o),3}^2/R_{s(o),2}^2$ with two $K_T$ bins.}
\label{HBT-Ratio}
\end{figure*}

Tables~\ref{tab:HBTRFitPars-1} and~\ref{tab:HBTRFitPars-2} show the fit parameters $R_{s(o),n}^2$ at different hadronic rescattering time $t_{hs}$ for three initial $^{12}$C configurations.
The amplitude of $R_{s(o),0}^2$ takes the similar value for chain structure and triangle structure, while the amplitudes for Woods-Saxon distribution case is slightly smaller than the others.
$R_{s,0}^2$ increases with the increase of $t_{hs}$ for all configurations of $^{12}$C, while $R_{s,2}^2$ decreases.
This $t_{hs}$ dependence of $R_{s,2}^2$ and $R_{s,0}^2$ reflects the approximate description of the eccentricity of the colliding system~\cite{Plumberg:2013nga,Bozek:2014hwa}, as the final eccentricity $\epsilon_{2}^f \approx 2R_{s,2}^2/R_{s,0}^2$.
It is noted that the dependence of the source eccentricity on the lifetime of the system has been investigated~\cite{Lisa-2}.
From the tables, it is obvious that $\epsilon_2^f$ for chain structure is larger than that for other cases, although the system undergoes hadronic rescattering.
$R_{o,0}^2$ and $|R_{o,2}^2|$ more or less increase with the increase of $t_{hs}$, while $|R_{o,2}^2|$ with high $K_T$ does not present an obvious $t_{hs}$ dependence.
$R_{s,3}^2$ and $|R_{o,3}^2|$ for three configurations of $^{12}$C also have an increasing tendency with $t_{hs}$, and triangle structure always has the largest absolute value.
Even though the third-order harmonic eccentricity coefficient $\epsilon_3^f$ cannot be obtained directly as $\epsilon_2^f \approx 2R_{s,2}^2/R_{s,0}^2$ as discussed in reference~\cite{Plumberg:2013nga}, the value of $R_{s(o),3}^2$ for triangle structure can still reflect the initial geometry effect on final state.
Since chain (triangle) structure always has the largest second (third)-order oscillation, the ratio of the third-order to the second-order coefficient $R_{s(o),3}^2/R_{s(o),2}^2$ can differentiate these three configurations as shown in Fig.~\ref{HBT-Ratio}.
And the effect of different configurations is clearer and more stable with high $K_T$.
For example, with high $K_T$, $R_{o,3}^2/R_{o,2}^2$ for Woods-Saxon distribution is always around 0.5.
For chain structure, it is no more than 0.2 although finally it gets larger than the value at the beginning.
For triangle structure, it is always larger than 0.6 and even up to 1.
The reason may be that pions with high $p_T$ are more likely to come from the center of the source at earlier stage.
Other results of $|R_{o(s),3}^2/R_{o(s),2}^2|$ for the three cases also show a consistent order, which from large to small is triangle, Woods-Saxon, chain.

Furthermore, we investigated the eccentricity of the system at the initial stage, which is calculated by~\cite{Alver:2010gr},
\begin{equation}
\epsilon_n^i = \frac{\sqrt{\langle r^2\cos(n\phi)\rangle^2 + \langle r^2\sin(n\phi)\rangle^2}}{\langle r^2\rangle},
\label{eccentricity}
\end{equation}
where $r$ and $\phi$ are the polar coordinate positions in transverse plane of participant nucleons.
Table~\ref{tab:eccent} presents the final eccentricity estimated by $2R_{s,2}^2/R_{s,0}^2$ and the initial eccentricity for different configurations of $^{12}$C.
It is expected that the second-order eccentricity takes the largest value for chain structure and the third-order eccentricity is larger for triangle structure than that for other cases as $|R_{s(o),3}^2|$ listed in Table~\ref{tab:HBTRFitPars-1} and~\ref{tab:HBTRFitPars-2}.
And the final eccentricity estimated with high $K_T$ is closer to the initial eccentricity.
These results of $R_{s(o),n}^2$ (n=2,3) indicate the initial geometry can be reflected in the final observables sensitive to azimuthal angle such as HBT correlation and collective flow~\cite{Zhang:2017xda,Zhang:2018,bozek2014alpha,broniowski2014signatures}.
Further, the system scan experiments around $^{12}$C may provide the platform to examine if there are $\alpha$ clusters in the carbon nuclei and to distinguish the structure by comparing the HBT correlation and collective flow between $^{12}$C+A collisions and other colliding systems.

\begin{table}[H]
\caption{ \label{tab:eccent} The final eccentricity compared with the initial eccentricity for different configurations of $^{12}$C.}
\begin{ruledtabular}
\begin{tabular}{llll}
 & Woods-Saxon & Triangle & Chain\\\hline
$2R_{s,2}^2/R_{s,0}^2$(low $K_T$) & 0.0458 & 0.0467 & 0.0979\\
$2R_{s,2}^2/R_{s,0}^2$(high $K_T$) & 0.147 & 0.138 & 0.297\\
$\epsilon_2^i$ & 0.263 & 0.227 & 0.499\\
$\epsilon_3^i$ & 0.219 & 0.296 & 0.179
\end{tabular}
\end{ruledtabular}
\end{table}

\section{Summary}
\label{SecSummary}

Configuration of $^{12}$C is initialized as $\alpha$-clustered triangle, $\alpha$-clustered chain and Woods-Saxon distribution of nucleons, and then the $^{12}$C + $^{197}$Au central collisions at $\sqrt{s_{NN}} =$ 200 GeV are simulated by the string melting AMPT model.
The azimuthal dependences of the HBT radii relative to the second- and third-order participant plane from $\pi$-$\pi$ correlation are presented with different hadronic rescattering time, showing the possibility as a probe of different initial geometries.
For quantitative description of the oscillation of the HBT radii and the system eccentricity, the final eccentricity were estimated by the zeroth- and second-order HBT radii, and compared with the initial eccentricity.
According to our calculations, the ratio of the third-order to the second-order HBT radii is quite sensitive to different configurations of $^{12}$C.
And we suggest performing azimuthally differential femtoscopy with high $p_T$ identical particles.
From the results, the azimuthal angle dependence of the HBT radii relative to the second- and third-order participant plane can be taken as an effective probe to distinguish the exotic nuclear structure besides collective flow.

\begin{acknowledgments}

This work was supported in part by the National Natural Science Foundation of China under contract Nos. 11890714, 11875066, 11421505, and 11775288, National Key R\&D Program of China under Grant No. 2016YFE0100900 and 2018YFE0104600, the Key Research Program of Frontier Sciences of the CAS under Grant No. QYZDJ-SSW-SLH002, and the Key Research Program of the CAS under Grant NO. XDPB09.

\end{acknowledgments}

\bibliographystyle{unsrtnat}
\bibliography{HBT_arxiv}
%\end{CJK*}
\end{document}